\documentclass{ws-procs9x6}

\newcommand{\inter}{{\mbox{\scriptsize int}}}

\newcommand{\bG}{\bar{G}}
\newcommand{\bV}{\bar{V}}

\newcommand{\bSigma}{\bar{\Sigma}}

\newcommand{\bLambda}{\bar{\Lambda}}

\begin{document}

\title{Renormalization and gauge symmetry\\ for 2PI effective actions\footnote{\uppercase{W}ork in collaboration with \uppercase{J}. \uppercase{B}erges, \uppercase{S}. \uppercase{B}ors\'anyi \& \uppercase{J}. \uppercase{S}erreau from the \uppercase{I}nstitut for \uppercase{T}heoretical \uppercase{P}hysics, \uppercase{U}niversity of \uppercase{H}eidelberg, \uppercase{G}ermany.}}

\author{U. Reinosa}

\address{Institut f\"ur Theoretische Physik, Technische Universit\"at Wien,\\
Wiedner Hauptstrasse 8-10/136, A-1040 Wien, Austria}

\maketitle

\abstracts{We explore a method to recover symmetry identities in the 2PI formalism. It is based on non-perturbative approximations to the 1PI effective action. We discuss renormalization questions raised by this technique.}

\section{Introduction}
In the 2PI resummation scheme, the quantum fluctuations encoded in the effective action are expressed in terms of 1- and 2- point fields $\phi$ and $G$ via the so-called CJT loop expansion\cite{CJT}: $\Gamma[\phi,G]=S_0[\phi]+(i/2)\mbox{Tr }\ln\,G^{-1}+(i/2)\mbox{Tr }G\,G_0^{-1}-i\Gamma_\inter[\phi,G]$, where $S_0[\phi]$ denotes the free, classical action and $\Gamma_\inter[\phi,G]$ is made of two-particle irreducible diagrams (2PI) with vertices arising from the shifted theory $S[\phi+\varphi]-S[\phi]-\varphi\,\cdot\,\delta S/\delta\varphi|_{\varphi=\phi}$ and propagator $G$. The physical point of the system corresponds to the stationnarity point of $\Gamma$. Such a formulation of the quantum theory is useful in many situations, from describing the thermodynamics of the equilibrated quark-gluon plasma\cite{BIR} to describing far from equilibrium dynamics and equilibration of quantum fields\cite{Berges}. However this formalism presents important difficulties in the case of gauge theories where $G$ does not fulfill usual Ward-Takahashi identities. We explore here a way to define $n$-point functions satisfying the usual Ward-Takahashi identities, in particular a transverse propagator. This approach has already been followed in the case of the $O(N)$ model\cite{vanHeesKnoll2} and allows to recover the Goldstone theorem in the case of a spontaneously broken symmetry. We apply it here to abelian gauge theories. We furthermore complete the scalar analysis\cite{vanHeesKnoll2} by explaining how to renormalize the scheme.

\section{Symmetries}
The idea is to construct $n$-point functions $\Gamma^{(n)}$ from the derivatives of $\Gamma[\phi]\equiv\Gamma[\phi,\bG[\phi]]$ that is the 2PI effective action evaluated at the stationnary propagator in presence of a mean field value $\phi$. If both definitions $\bG$ and $\Gamma^{(2)}$ for the propagator agree in the exact theory, this is not the case when one considers approximations. $\bG$ is then interpreted as a mere tool to build up resummations whereas $\Gamma^{(2)}$ is a more reliable object as it satisfies usual symmetry identities, at least in the $O(N)$ model\cite{vanHeesKnoll2} or in QED.

It is worth noting here that the symmetry is still present at the level of the 2PI effective action. In the case of QED, we thus have a 2PI Ward-Takahashi identity\cite{BBRS}, even within a truncation. From this one can infer a symmetry property for the effective action $\Gamma[\mathcal{A}_{\mu},\bar\Psi,\Psi]$ constructed as explained in the previous paragraph. More precisely, one can show that the effective action:
\begin{equation}
\Gamma^{*}[\mathcal{A}_{\mu},\bar\Psi,\Psi]=\Gamma[\mathcal{A}_{\mu},\bar\Psi,\Psi,\bar{G}_{\mu\nu},\bar{S}]-\frac{1}{2}\mathcal{A}^\nu\left\{\mu^2+\frac{1}{\lambda}\partial^2\right\}\mathcal{A}_\nu\,,
\end{equation}
is invariant under a usual gauge transformation\footnote{This can be directly checked on the CJT formula using the transformation of the auxiliary propagators: $\delta \bG=0\,,\quad\delta \bar{S}(x,y)=i\Big\{\Lambda(x)-\Lambda(y)\Big\}\bar{S}(x,y)$.} ($\mu$ and $\lambda$ denote respectively an infrared regulator for the photon and the gauge fixing parameter in the covariant gauge; $\bar{G}_{\mu\nu}$ and $\bar{S}$ denote the stationnary propagators in presence of mean fields). From this effective action it is then easy to construct transverse $n$-point functions, in particular a trustworthy propagator.

\section{Renormalization}
The aim of this section is to show how to renormalize the functional $\Gamma[\phi]$ constructed as explained in the previous section. We will concentrate here in the case of a $4$--dimensional $Z_2$--symmetric scalar theory with $\varphi^4$ interaction and leave the inclusion of fermions and gauge symmetry for subsequent publications\cite{BBRS}. The functional $\Gamma[\phi]$ is nothing but the 1PI effective action which is known to be renormalizable in a 1PI loop expansion\cite{ZinnJustin}. However, in the present case, the functional is constructed from approximations at the level of the 2PI effective action, and thus renormalization has to be revisited. We will not give here a complete proof of renormalizability\cite{BBRS} but rather insist on a few points which deserve special attention.

\begin{figure}[htbp]
\begin{center}
\includegraphics[width=10cm]{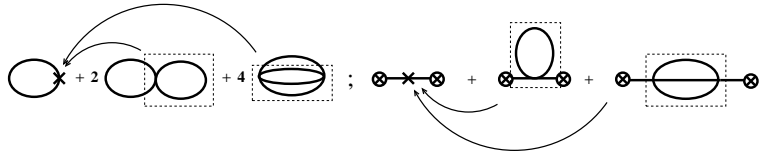}
\caption{Renormalization of 2-point singularities using standard BPH techniques on resummed diagrams. Only two diagrams absorb these singularities.\label{fig:ex}}
\end{center}
\end{figure}

Part of the discussion can be followed directly on the CJT\cite{CJT} formula for the 2PI effective action. The analysis of divergences is based on standard BPH\cite{BPH} techniques applied to diagrams with resummed propagators. For such an analysis, it is convenient to use a field expansion of the effective action: $\Gamma_\inter[\phi,G]=\sum_{n=0}^\infty\Gamma_{2n}[G]\,\phi^{2n}$. For a given $n$, the approximation for $\Gamma_{2n}[G]$ consists of a certain number of diagrams with $2n$ external legs. The strategy is then to draw boxes in order to track divergences: 2-point boxes represent mass and field strength singularities whereas 4-point boxes represent coupling singularities. Such divergences are absorbed in diagrams with lower number of loops, namely the topologies obtained by shrinking BPH boxes into points \footnote{\uppercase{T}hus not all the approximations are renormalizable. For an approximation to be renormalizable, it has to be such that whenever one includes a diagram, one has also to include all the diagrams related to it by the BPH procedure. In general we will use a 2PI loop expansion which is renormalizable.}. As far as $2$-point singularities are concerned, the result is quite simple: The diagrams in the effective action being 2PI, the only $2$-point boxes one can draw are either boxes on $0$ legs diagrams, with one of the lines of the diagram lying outside the box, or boxes drawn on two legs diagrams, encapsulating the whole diagram. These two situations correspond to the two cross diagrams in Fig. \ref{fig:ex} which are the only ones carrying mass and field strength counterterms in the CJT formula.

\begin{figure}[htbp]
\begin{center}
\includegraphics[width=4cm]{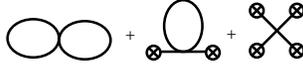}
\caption{Diagrams related to the renormalization of divergences arising from the non-perturbative structure of the propagator. They are all local.\label{fig:tadpoles}}
\end{center}
\end{figure}

The previous analysis misses part of the $4$-point (coupling) singularities related to the non-perturbative nature of the propagator. These divergences emerge from an overlap between the logarithmic behaviour of the propagator and that of the 2PI diagram under study and call for analysing the diagrammatic content of the resummed propagator $\bG$.\cite{BIU,vanHeesKnoll} In the present case, to easily disentangle such divergences, it is much more convenient to work with proper vertices, that is, derivatives of the effective action. We work here in the symmetric phase but the very same counterterms can be used in the broken phase\cite{BBRS}. The structure of proper vertices is easily obtained by taking derivatives of the equation (which is itself derived from the CJT formula and the stationarity condition):
\begin{eqnarray}
\label{eq:secder}
 i\frac{\delta^2 \Gamma}{\delta\phi_1\delta\phi_2} &=& 
 -G_{0,12}^{-1}+\frac{\partial^2\Gamma_\inter}{\partial\phi_1\partial\phi_2}
 +\frac{\partial^2\Gamma_\inter}{\partial\phi_1\partial G_{ab}}\bG_{ac}\bG_{bd}
 \frac{\delta \bSigma_{bd}}{\delta \phi_2}\,,
\end{eqnarray}
where $\partial$ and $\delta$ denote partial and total derivatives respectively. Thus proper vertices will consist of loop diagrams involving partial derivatives of $\Gamma_\inter$ with respect to $\phi$ or $G$ (which we call 2PI kernels) and total derivatives of $\bSigma$, connected via the resummed propagator $\bG$. For example in the symmetric phase, the $2$-point function consists of the first two terms on the r.h.s. of Eq. (\ref{eq:secder}) whereas the $4$-point proper vertex reads: 
\begin{eqnarray}
\label{eq:phi4_4pf1}
 i\Gamma^{(4)}_{1234}&=& 
 \frac{\partial^4\Gamma_\inter}{\partial\phi_1\cdots\partial\phi_4}
 +\left(\frac{\partial^3\Gamma_\inter}{\partial\phi_1\partial\phi_2\partial G_{ab}}
 \,\bG_{ac}\bG_{bd}\,\frac{\delta^2\bSigma_{bd}}{\delta\phi_3\delta\phi_4}
 +{\rm perm.}\right)\,,
\end{eqnarray}
where 'perm.' denotes the two permutations $(12,34)\rightarrow(13,24)$ and 
$(12,34)\rightarrow(14,23)$ of the term between brackets. The total derivative 
appearing on the r.h.s. satisfies the following integral equation:
\begin{equation}
\label{eq:phi4_4pf2}
 \frac{\delta^2\bSigma_{ab}}{\delta\phi_1\delta\phi_2}=
 \left(2\frac{\partial^3\Gamma_\inter}
 {\partial G_{ab}\partial\phi_1\partial\phi_2}\right)
 +\left(2\frac{\partial^2\Gamma_\inter}{\partial G_{ab}\partial G_{cd}}\right)\,
 \bG_{ce}\bG_{df}\,\frac{\delta^2\bSigma_{ef}}{\delta\phi_1\delta\phi_2}\,.
\end{equation}
Renormalization of $\Gamma[\phi]$ then amounts to first renormalizing the 2PI kernels, next total derivatives of $\bSigma$ and finally the proper vertices. The BPH approach described previously renormalizes almost all the kernels. Only two kernels deserve much more attention due to the non-perturbative character of the propagator. These kernels are the 2-point kernels $\partial\Gamma_\inter/\partial G$ and $\partial^2\Gamma_\inter/\partial\phi^2$. $\partial\Gamma_\inter/\partial G$ has been extensively discussed in Refs.\cite{BIU,vanHeesKnoll}. There it was shown that the coupling singularities hidden in the kernel are exactly accounted for by a Bethe-Salpeter equation for a 4-point auxiliary function $\bV=\bLambda+\bLambda\,\bG^2\,\bV=\bLambda+\bV\,\bG^2\,\bLambda$, where $\bLambda=2\partial^2\Gamma_\inter/\partial G^2$. Renormalization of $\bV$ is performed via a shift of the tree level (local) contribution to the kernel $\bLambda$. At the level of the effective action, this contribution corresponds to the ``eight'' diagram depicted in Fig. \ref{fig:tadpoles}. The same results apply to the kernel $\partial^2\Gamma_\inter/\partial\phi^2$. Coupling singularities are enterily described by a second Bethe-Salpeter equation for $V$ (this equation can be solved in terms of $\bV$): $V=\Lambda+V\,\bG^2\,\bLambda=\Lambda+\Lambda\,\bG^2\,\bV$, with $\Lambda=2\partial^3\Gamma_\inter/\partial G\partial\phi^2$. Renormalization amounts to a shift of the tree level (local) contribution to the kernel $\Lambda$. At the level of the effective action, this contribution corresponds to the ``tadpole'' diagram depicted in Fig. \ref{fig:tadpoles}
(let us note here that self-consistency and the 2PI character of the resummation are crucial for the renormalization program to work at all\cite{BIU,BBRS}). Indeed, this last result allows, in the symmetric phase, to also renormalize the $2$-point function which is nothing but the kernel $\partial^2\Gamma_\inter/\partial\phi^2$. Renormalization of the $4$-point function is not so straightforward as it involves the quantity $\delta^2\bSigma/\delta\phi^2$. However the latter is nothing but $V$ as it satisfies the same self-consistent equation (see Eq. (\ref{eq:phi4_4pf2})). Thus $\delta^2\bSigma/\delta\phi^2$ is finite. We can now analyze divergences in the $4$-point function. To this purpose, it is convenient to express $\delta^2\bSigma/\delta\phi^2=V$ in terms of $\bV$: $i\Gamma^{(4)}=\partial^4\Gamma_\inter/\partial\phi^4+2\Big(\Lambda\,\bG^2\,\framebox{$\Lambda$}+\Lambda\,\bG^2\,\framebox{$\bV\,\bG^2\,\Lambda$}+{\rm perm.}\Big)$. The boxes drawn in the previous equation show that all the subdivergences are indeed already taken into account by the renormalization of $V$. There only remain overall divergences which are absorbed in a shift of the tree level contribution (local) to $\partial^4\Gamma_\inter/\partial\phi^4$. At the level of the effective action, this contribution arises from the third diagram depicted in Fig. \ref{fig:tadpoles}.

Once all the kernels have been renormalized together with $\bV$ and $V$, renormalization of total derivatives of $\bSigma$, $\delta^p\bSigma/\delta\phi^p$ (with $p>2$) and proper vertices $\Gamma^{(n)}$ (with $n>4$) is straightforward. Indeed one can show\cite{BBRS} that it is possible to express the latter in terms of diagrams made of finite kernels, or Bethe-Salpeter functions $\bV$ and $V$, and only involving finite loops.

To extend the results on renormalization to QED\cite{BBRS}, one first needs to incorporate fermions. Next one needs to discuss spurious UV singularities in the auxiliary photon propagator $\bG_{\mu\nu}$\cite{BBRS}.



\begin{thebibliography}{0}
\bibitem{CJT} J. M. Cornwall, R. Jackiw and E. Tomboulis, Phys. Rev. D {\bf 10} (1974).

\bibitem{BIR} J.-P. Blaizot, E. Iancu and Anton Rebhan, Phys. Rev. D {\bf 63} (2001).

\bibitem{Berges} J. Berges, Nucl. Phys A {\bf 699} (2002). J. Berges, S. Bors\'anyi and J. Serreau, Nucl. Phys. B {\bf 660} (2003).

\bibitem{vanHeesKnoll2} H. van Hees and J. Knoll, Phys. Rev. D {\bf 66} (2002).

\bibitem{BBRS} J. Berges, S. Bors\'anyi, U. Reinosa and J. Serreau, work in progress.

\bibitem{ZinnJustin} J. Zinn-Justin, Quantum Field Theory and Critical Phenomena, Oxford Science Publications.

\bibitem{BPH} N. N. Bogoliubov, O. S. Parasiuk, Acta Math. {\bf 97} (1957). K. Hepp, Commun. Math. Phys. {\bf 2} (1966).

\bibitem{BIU} J.-P. Blaizot, E. Iancu and Urko Reinosa, Phys. Lett. B {\bf 568} (2003), Nucl. Phys. A {\bf 736} (2004).

\bibitem{vanHeesKnoll} H. van Hees and J. Knoll, Phys. Rev. D {\bf 65} (2002).

\end{thebibliography}
\end{document}